\documentclass[10pt,conference]{IEEEtran}
\IEEEoverridecommandlockouts{}
\usepackage{flushend}
\usepackage{algorithmic}
\usepackage{graphicx}
\usepackage{textcomp}
\usepackage{xcolor}
\usepackage[normalem]{ulem}
\usepackage[hidelinks]{hyperref}
\usepackage[protrusion=true,expansion=true,tracking=false,kerning=true,spacing=true]{microtype}

\microtypecontext{spacing=nonfrench}
\usepackage{tcolorbox}
\usepackage{ifthen}
\usepackage{flushend}
\usepackage{graphicx}
\usepackage{tcolorbox}
\usepackage{booktabs}

\newtcolorbox{custombox}[1]{
	colback=gray!10,
	colframe=gray!70,
	left=1.5mm,
	right=1.5mm,
	top=1.5mm,
	bottom=1.5mm,
	fonttitle=\bfseries,
	arc=0mm,
	leftrule=1mm,
	rightrule=0mm,
	toprule=0mm,
	bottomrule=0mm,
	notitle,
	before=\par\noindent,
	before upper={\textbf{#1: } },
}

\usepackage{xspace}
\usepackage{pifont}
\usepackage{fontawesome}
\usepackage{tikz}
\usepackage{pgfplots}
\usetikzlibrary{intersections}
\usepgfplotslibrary{fillbetween}
\pgfplotsset{compat=1.12}
\usepackage{multirow}
\usepackage{xspace}

\usepackage{listings}
\usepackage{xcolor}

\usepackage{ragged2e} 

\definecolor{codeblue}{rgb}{0.22, 0.45, 0.70}
\definecolor{codegreen}{rgb}{0.0, 0.5, 0.0}
\definecolor{codered}{rgb}{0.6, 0, 0}
\definecolor{codegray}{rgb}{0.5, 0.5, 0.5}

\lstdefinestyle{java}{
  language=Java,
  basicstyle=\ttfamily\footnotesize,
  keywordstyle=\color{codeblue}\bfseries,
  stringstyle=\color{codered},
commentstyle=\color{codegreen}\itshape,
  numbers=left,
  numberstyle=\tiny\color{codegray},
  stepnumber=1,
  numbersep=10pt,
  backgroundcolor=\color{white},
  showspaces=false,
  showstringspaces=false,
  tabsize=4,
  breaklines=true,
  breakatwhitespace=true,
  frame=single,
  rulecolor=\color{codegray},
  captionpos=b,
  escapeinside={(*@}{@*)},
      abovecaptionskip=10pt
}

\lstdefinestyle{console}{
basicstyle=\ttfamily\footnotesize, 
    backgroundcolor=\color{gray!20}, 
    frame=single,
        frame=none, 
    breaklines=true,
    keepspaces=true,
    columns=fixed,
    deletekeywords={new, for},
    abovecaptionskip=10pt,
        numbers=none 
}

\usepackage{float}
\usepackage{placeins}
\usepackage{fancybox,framed}
\usepackage[utf8]{inputenc}

\usepackage[figure,norelsize,vlined,linesnumbered]{algorithm2e}
\SetAlFnt{\small}
\SetAlCapFnt{\large}
\SetAlCapNameFnt{\large}

\SetCommentSty{mycommfont}

\usepackage{booktabs} 
\usepackage{xspace}
\usepackage{tabularx,colortbl}
\usepackage{multirow, makecell}
\usepackage{paralist}
\usepackage{tabularx}
\usepackage{listings}
\usepackage{graphicx}
\usepackage{ifthen}
\usepackage{amsthm}
\usepackage{indentfirst}
\usepackage{xcolor}

\usepackage{pifont}
\definecolor{darkjunglegreen}{rgb}{0.000000,0.392157,0.000000}

\usepackage{url}
\def\codesize{\small}
\def\<#1>{\codeid{#1}}\protected\def\codeid#1{\ifmmode{\mbox{\codesize\ttfamily
			#1}}\else{\codesize\ttfamily
		#1}\fi}

\definecolor{mygray}{gray}{0.9}
\usepackage{fancybox}
\usepackage{listings}
\usepackage{etoolbox}
\usepackage{kvoptions}
\makeatletter
\newcommand{\removelatexerror}{\let\@latex@error\@gobble}
 
\usepackage{xspace}

\newcommand{\starcoder }{\textsc{StarCoder}\xspace}
\newcommand{\gpt }{\textsc{GPT-4o}\xspace}

\newtcolorbox{findings}{
    left,
    width=\linewidth,
    halign=left,
    colframe=gray,
    boxrule=1pt,
    leftrule=2pt,
    bottomrule=2pt,
    left=5pt,
    right=5pt
}

\SetKwProg{Fn}{Function}{}{}
\makeatother

\usepackage{booktabs} 
\usepackage{graphics} 
\usepackage{multirow,bigstrut}
\usepackage[framemethod=tikz]{mdframed}
\usepackage{graphicx}
\usepackage{bm} 
\usepackage{relsize}

\makeatletter
\DeclareRobustCommand{\etc}{%
    \@ifnextchar{.}%
        {etc}%
        {etc.\@\xspace}%
}

\usepackage{bold-extra}

\DeclareMathAlphabet\mathbfcal{OMS}{cmsy}{b}{n}

\usepackage{multirow,array,varwidth}
\usepackage{booktabs}
\usepackage{pifont}

\newcommand{\mytodoblue}[1]{\textcolor{blue}{\ding{46}~{\sf}~#1}}

\usepackage{listings}
\usepackage{color,url}

\definecolor{javared}{rgb}{0.6,0,0} 
\definecolor{javagreen}{rgb}{0.25,0.5,0.35} 
\definecolor{javapurple}{rgb}{0.5,0,0.35} 
\definecolor{javadocblue}{rgb}{0.25,0.35,0.75} 
\usepackage{lipsum}
\usepackage{paralist}

\newtheoremstyle{mystyle}
{0.0mm}
{0.0mm}
{}
{}
{\bfseries}
{.}
{ }
{}

\usepackage{amsbsy}
\definecolor{orange}{rgb}{1,0.5,0}
\definecolor{darkjunglegreen}{rgb}{0.000000,0.392157,0.000000}

\usepackage{soul}

\newcommand{\gunel}[1]{\mytodoblue{[Gunel: #1]}}

\usepackage{paralist}

\font\tt=rm-lmtl10

\usepackage{enumitem}
\usepackage{amsmath}
\usepackage{microtype}
\usepackage{amsthm}

\newlength\marincrease


\AtBeginDocument{%
  }

\author{
    \IEEEauthorblockN{Adam Bodicoat\IEEEauthorrefmark{1}, Gunel Jahangirova\IEEEauthorrefmark{2}, and Valerio Terragni\IEEEauthorrefmark{1}}
    \IEEEauthorblockA{\IEEEauthorrefmark{1}University of Auckland, Auckland, New Zealand\\
    Email: abod278@aucklanduni.ac.nz, v.terragni@auckland.ac.nz}
    \IEEEauthorblockA{\IEEEauthorrefmark{2}King's College London, London, United Kingdom\\
    Email: gunel.jahangirova@kcl.ac.uk}
}

\title{Understanding LLM-Driven Test Oracle Generation}

\begin{document}
\maketitle

\IEEEpubid{\begin{minipage}{\textwidth}
\vspace{16mm}
\centering \large This is the authors’ version of the paper that has been accepted for publication in the \\ 2nd ACM/IEEE International Conference on AI‑Powered Software (AIware 2025)
\end{minipage}}

\begin{abstract}
 Automated unit test generation aims to improve software quality while reducing the time and effort required for creating tests manually. However, existing techniques primarily generate regression oracles that predicate on the implemented behavior of the class under test. They do not address the oracle problem: the challenge of distinguishing correct from incorrect program behavior. 
 
 With the rise of Foundation Models (FMs), particularly Large Language Models (LLMs), there is a new opportunity to generate test oracles that reflect intended behavior. This positions LLMs as enablers of Promptware, where software creation and testing are driven by natural-language prompts.
 
 This paper presents an empirical study on the effectiveness of LLMs in generating test oracles that expose software failures. We investigate how different prompting strategies and levels of contextual input impact the quality of LLM-generated oracles. Our findings offer insights into the strengths and limitations of LLM-based oracle generation in the FM era, improving our understanding of their capabilities and fostering future research in this area.

\end{abstract}





\section{Introduction}

Software testing is crucial for ensuring software quality and reliability. However, manually creating test oracles is labor-intensive. While automated test generation tools like \textsc{EvoSuite}~\cite{fraser2011evolutionary} and \textsc{Randoop}~\cite{pacheco2007randoop} typically rely on regression oracles, which assume the current version is correct~\cite{Ruberto2025FromImplemented,konstantinou2024llms}. This limits their ability to detect faults in the given version, this is the long-standing \textbf{oracle problem} in test automation~\cite{barr2014oracle,terragni-fse-2020,Ruberto2025FromImplemented}.

In the era of \textbf{Foundation Models (FMs)}, Large Language Models (LLMs) offer new opportunities to address this problem~\cite{terragni2025future}. With their capabilities in natural language understanding, pattern recognition, and contextual reasoning, LLMs can help bridge the gap between automated test generation and meaningful oracle creation~\cite{jiang2024survey, yang2024evaluation,ryan2024code,hossain2024doc2oracle,augmentest}. By interpreting developer intent through prompts, LLMs can generate oracles that align with expected behaviors rather than implemented ones~\cite{konstantinou2024llms, binta2024togll,Ruberto2025FromImplemented,jahangirova2023sbfttrack}.

Recent research has started to explore LLMs for generating tests and oracles~\cite{binta2024togll, konstantinou2024llms, hayet2024c, yuan2023no, yang2024evaluation,ravi2025llmloop}, but critical questions remain. In particular, regarding how prompting strategies and input context affect oracle quality. Understanding these factors is essential as we transition toward \textbf{Promptware}, where software validation may be driven by natural-language prompts crafted by both developers and non-developer prompt experts.

To address this gap, this paper present an empirical study on how prompting techniques (zero-shot, few-shot, chain-of-thought [CoT], and tree-of-thoughts [ToT]) and contextual inputs (test prefix alone, test prefix plus method under test [MUT], and test prefix plus class under test [CUT]) influence the quality of LLM-generated test oracles.  Our study isolates the oracle-generation capability of LLMs rather than the broader task of fault reproduction. We therefore do not include failure descriptions or bug reports in the prompt. Incorporating such inputs would shift the task toward bug reproduction or localization, which are orthogonal to the oracle problem. To the best of our knowledge, this is the first study to conduct such an analysis.

We used the \textbf{GitHub Recent Bugs (GHRB)} benchmark~\cite{lee2024github}, a dataset designed to evaluate LLMs on real-world Java bugs while mitigating data leakage risks. For each bug, we provide to LLMs the buggy code and its triggering test input -- omitting the original oracle -- and analyze oracle quality based on compilation success and bug exposure.
Our experiments involved 36 representative GHRB bugs and two LLMs, \gpt and \starcoder.

Our results reveal several important trends:

\ding{172} Oracles generated with more context compile and detect bugs more reliably. CUT-level context significantly outperforms other configurations, achieving 53.64\% accuracy versus 40.74\% (MUT) and 40.38\% (test prefix only). This is an expected result.

\ding{173} Prompting style matters: zero-shot and few-shot prompts yield higher compilation rates (67.38\% and 72.96\%) and accuracy (54.56\% and 51.30\%) than CoT and ToT, which struggle with low compilation (both below 50\%).

\ding{174} Incorporating the CUT in the input prompt, along with zero-shot and few-shot prompting techniques, leads to the most consistently accurate LLM-generated test oracles. However, our findings show there is potential for reasoning based prompt techniques like CoT and ToT to be able to produce accurate test oracles given their high accuracy when they do produce compilable assertions.

\ding{175} While \starcoder slightly outperforms \gpt in average accuracy, \gpt paired with CUT context delivers the most consistently accurate oracles across all combinations.

 \ding{176} Prompting strategy has a stronger impact on oracle effectiveness than LLM choice.

These findings suggest that prompt design and context play a critical role in the effectiveness of LLM-based oracle generation. While reasoning-driven prompting (e.g., CoT, ToT) shows potential when it compiles, zero-shot and few-shot prompting currently offer the best tradeoff between accuracy and robustness. Our study offers guidance for AI-assisted testing tools usable by both testing and prompt experts in the FM era.

In summary, this paper makes the following contributions:

\begin{itemize}
\item An empirical study of how prompt types and contextual inputs affect LLM-generated test oracle quality.
\item In support of Promptware, a series of insights into the effectiveness of LLMs in generating test oracles, providing guidance for the future of LLM-driven oracle generation.

\item Public release of code to ensure reproducibility and support future research in this area~\cite{data}.
\end{itemize}

\section{Experimental Design}
Our study aims to assess the ability of LLMs to generate accurate and correct test oracles using different input prompt variations. These variations fall into two categories: the content of the prompt and the prompting technique. Specifically, our empirical study investigates the following research questions:

\noindent \textbf{RQ1 - Input Context} What influence does the content and context within the prompt have on the accuracy of LLM generated test oracles?

\noindent \textbf{RQ2 - Prompt Engineering} How do different prompt engineering techniques impact the accuracy of LLM generated test oracles relative to each other?

\noindent \textbf{RQ3 - Model Comparison} How do different LLMs impact the accuracy of LLM generated test oracles?

\noindent \textbf{RQ4 - Impact Analysis} Which factor has the most significant impact on LLM-generated test oracles?

RQ1 studies how the content of the prompt influences LLM-generated test oracles. This helps determine what content provides the best context for test oracle generation. We use three levels of context: Test Prefix, Test Prefix with Method Under Test, and Test Prefix with Class Under Test.

RQ2 investigates the impact the prompting technique has on the accuracy of LLM generated test oracles. This allows us to compare the effectiveness of different prompting techniques in generating test oracles from existing test prefixes. In this study, we will be using four prompting techniques: zero shot, few shot, chain of thought, and tree of thoughts.

RQ3 examines how different LLMs impact test oracle generation based on their training differences.

RQ4 analyses the impacts the prompt content, prompting technique, and LLM have on the accuracy of the generated test oracles relative to each other. This allows us to determine the relative importance of each of these factors for LLM driven test oracle generation. In this study, we will measure the impact each variable we study has using the range of average accuracies for that variable. We also compare each result to find the factors which contribute to the highest average accuracy.

To answer these RQs, we conducted an experiment asking the selected LLMs to generate test oracles for each prompt, where the prompt is a unique combination of the prompt context and prompting technique. The test cases used fail on the buggy version of the code but pass on the corrected version. We remove the test oracle and provide the modified test case and the buggy code as input to the LLM. We then prompt the LLM to generate a suitable test oracle. We evaluate the correctness based on whether the generated oracle fails the buggy version and passes the correct version. This allows us to analyse how prompt content, prompting technique, and LLM influence oracle performance. 
\subsection{Dataset}
This study uses the \textbf{GitHub Recent Bugs (GHRB)}~\cite{lee2024github} benchmark, a dataset designed for evaluating the performance of LLMs on real world code. GHRB ensures bugs and fixes come from real repositories updated after the training period of popular LLMs, including \starcoder and \gpt. This avoids data leakage and model memorisation~\cite{lee2024github}. As of December 2024, GHRB contains 107 bugs from 16 popular open source Java repositories. We selected 36 bugs from this dataset, based on criteria designed to ensure experimental consistency and validity.

\noindent \textbf{1) Absence of compile errors:} Both the buggy and correct versions of the selected repositories were verified to be free of compile-time errors. This criterion was critical for eliminating confounding factors that could distort the evaluation of the generated test oracles. Ensuring compilability allowed the analysis to focus solely on the semantic correctness of the generated oracles. Compilation errors were likely due to incomplete fixes, missing dependencies, or build configuration issues present in certain repositories within the benchmark dataset.

\noindent
\textbf{2) Random sampling:} To minimise selection bias and enhance the generalisability of the findings, the subset of bugs are randomly sampled from those without compilation errors in the benchmark~\cite{kitchenham2002principles}. This approach ensures that the selected bugs represent a diverse and unbiased subset of the dataset~\cite{nagappan2005use}. We choose to select 36 bugs only due to the high computation cost required for larger subsets.

Similar to established bug benchmarks such as \textsc{Defects4J}~\cite{just2014defects4j}, each repository in the dataset includes a buggy version, where bug-revealing test cases fail, and a corresponding correct version, where the same test cases pass. We also ensure that the bug requires a test assertion to be exposed and does not result in unwanted exceptions, which makes the test fails without the need of assertions.


\medskip
Figures~\ref{figbug-test} and~\ref{fig:bug-impl}  illustrate a concrete example from the \textsc{JSoup} repository in the GHRB benchmark. Specifically, Figure~\ref{figbug-test} presents a bug-revealing test case designed to verify the correctness of the copy constructor in the \texttt{Safelist} class. The test case first creates an instance of the \texttt{Safelist} (\texttt{safelist1}), copies it into a second instance (\texttt{safelist2}), and subsequently modifies the original object by adding an additional attribute (\texttt{"invalidAttribute"}). The assertion (highlighted in violet) explicitly checks that this newly added attribute in the original object should \textit{not} be recognized as safe in the copied instance, confirming that the copy constructor correctly creates an independent copy.

However, the buggy implementation of the copy constructor (shown in Figure~\ref{fig:bug-impl}) fails to perform a deep copy. Instead, it simply reuses references to internal collections (\texttt{attributes}, \texttt{tagNames}, etc.), leading to unintended side effects. As a result, any subsequent changes made to the original instance incorrectly propagate to the copied instance, causing the test case to fail. The goal of our study is to investigate the key factors influencing the effectiveness of LLMs in generating test oracles that accurately detect software faults.

\begin{figure}[t]
    \centering
    \begin{minipage}{\linewidth}
       \begin{lstlisting}[language=Java]
@Test
public void testCopyConstructor_noSideEffectOnAttributes() {
    Safelist safelist1 = Safelist.none().addAttributes(TEST_TAG, TEST_ATTRIBUTE);
    Safelist safelist2 = new Safelist(safelist1);
    safelist1.addAttributes(TEST_TAG, "invalidAttribute");

    (*@\textcolor{violet}{assertFalse(safelist2.isSafeAttribute(TEST\_TAG\, null\, new Attribute("invalidAttribute"\, TEST\_VALUE)));}@*)
}
\end{lstlisting}
    \end{minipage}
    \vspace{-4mm}
    \caption{Bug-revealing test case exposing incorrect copy behavior in Jsoup's \texttt{Safelist} class (GHRB benchmark)}
    \label{figbug-test}
\end{figure}

\subsection{Prompt Construction}
This study evaluates the impact of four prompting techniques on test oracle generation: zero shot, few shot, chain of thought, and tree of thoughts. These were chosen for their ability to capture different dimensions of model reasoning and generalization~\cite{wang2024software}.

\textbf{Zero-Shot prompting (Z)} involves asking the model to perform a task without providing any prior examples or structured guidance~\cite{brown2020language}. For test oracle generation, this technique entails supplying the model with a task description to generate test oracles for the given test prefix. This approach evaluates the model’s inherent understanding of the task based solely on the prompt content and the model’s pre-training.

\textbf{Few-Shot prompting (F)} involves providing the model with a limited number of examples illustrating the task at hand~\cite{brown2020language,wei2022chain}. In this study, we use three test oracle examples from the same repository as the test prefix to balance prompt length with relevant context~\cite{brown2020language,wei2022chain}. Using examples from the same repository ensures consistency, and the same examples are used for all bugs in that repository to reduce variability and focus on prompt variations.

\textbf{Chain of Thought (CoT) prompting (Ch)} guides the model through a structured reasoning process~\cite{kojima2022large}. Following prior work~\cite{wei2022chain}, the CoT prompt for generating oracles includes the following steps:
\begin{itemize}
\item Identifying the purpose of the test prefix and code under test (if provided)
\item Determining the expected outcome of the code and test
\item Formulating the correct test oracles
\end{itemize}
This approach is designed to encourage logical reasoning to improve the accuracy of the generated oracles.

\textbf{Tree of Thoughts (ToT) prompting (Tr)} extends chain of thought reasoning by exploring multiple potential paths to solve the problem before converging on the most plausible solution~\cite{VellumAI2023}. As per previous work~\cite{yao2024tree}, in the context of test oracle generation, the prompt includes the following steps to construct a ToT prompt:
\begin{itemize}
    \item Root thought: Initial understanding of the test prefix and code under test (if provided).
    \item Branching: Exploring different paths of reasoning.
    \item Expansion: Developing each reasoning path further.
    \item Pruning: Removing invalid or redundant paths.
    \item Combinations: Combining insights from valid branches into a set of coherent and accurate test oracles.
\end{itemize}
This method leverages the model’s ability to consider diverse reasoning paths, potentially improving robustness in scenarios with complex or ambiguous prompts.

To further analyse the impact of different prompt configs, each prompting technique was tested with three levels of \textbf{input content}:

\textbf{Test Prefix (T)}: The minimal context consisting of the test code without the test oracles. Figure~\ref{figbug-test} is an example of a bug revealing test that is used in the prompt input but with the test oracle on line seven removed.

\begin{figure}[t]
    \centering
    \begin{minipage}{\linewidth}
\begin{lstlisting}[language=Java]
/**
 Deep copy an existing Safelist to a new Safelist.
 @param copy the Safelist to copy
*/
public Safelist(Safelist copy) {
    this();
    tagNames.addAll(copy.tagNames);
    attributes.putAll(copy.attributes);
    enforcedAttributes.putAll(copy.enforcedAttributes);
    protocols.putAll(copy.protocols);
    preserveRelativeLinks = copy.preserveRelativeLinks;
}
\end{lstlisting}
    \end{minipage}
 \vspace{-4mm}
 \caption{Buggy implementation of the \texttt{Safelist} copy constructor causing unintended side-effects (GHRB benchmark)}
    \label{fig:bug-impl}
\end{figure}

\textbf{Test Prefix + MUT (M) }: Adds the method being tested to provide additional context for the oracle generation. Figure~\ref{fig:bug-impl} is an example of the method under test to be added to the prompt input corresponding to the test case in Figure~\ref{figbug-test}.

\textbf{Test Prefix + CUT (C)}: Adds the class containing the test and method, providing more comprehensive context.

In the prompts, all test assertion oracles are removed from the test prefix, which may contain multiple assertions. The prompt let the LLM decide how many assertions should be generated. The provided CUT and MUT come from the buggy version, but we do not indicate in the prompt that the test exposes a bug. This reflects a realistic scenario where the goal is to generate an oracle without prior knowledge of whether the test reveals a bug. We refine the prompt iteratively, following established prompt engineering guidelines and prior work on LLM-driven test oracle generation~\cite{binta2024togll,hayet2024c,konstantinou2024llms,yuan2023no,yang2024evaluation,zhangexploring}. We manually identified MUT and CUT by examining the code changes in the corresponding corrected version in the GHRB repository.
\emph{Due to space constraints, prompts are omitted but available in the supplementary material~\cite{data}}

\subsection{Models}
For this study, we selected two LLMs: \starcoder and \gpt, enabling comparison between a domain-specific and a general-purpose model regarding prompt effects on test oracle generation. These models were chosen for their popularity and because the GHRB benchmark avoids data leakage in them~\cite{lee2024github}.

\textbf{\starcoder (S)} (v. 1.0) is a domain-specific LLM for code completion, trained on a large code corpus as part of the BigCode Project~\cite{li2023starcoder}. Its focus on code generation makes it well suited for generating test oracles from test prefixes.

\textbf{\gpt (G)} (\texttt{gpt-4o-mini-2024-07-18}) is a general-purpose LLM trained on diverse data across domains (including source code)~\cite{openai2024gpt4technicalreport}, enabling it to handle tasks like code generation.

The inclusion of \gpt allows for an exploration of how prompt input influences oracle generation in a general-purpose model, providing a useful benchmark for comparing the performance of specialised and non-specialised models.

While other advanced reasoning-oriented models, such as \textsc{GPT-4o1}, are available, we intentionally selected \starcoder and \gpt due to their documented effectiveness in code-related tasks~\cite{li2023starcoder}. Although including reasoning-focused LLMs might offer additional insights, it remains uncertain whether using these newer models is feasible. The GHRB benchmark was specifically designed to prevent data leakage in \starcoder and \gpt, but it does not explicitly account for more recent models~\cite{lee2024github}. Moreover, \starcoder is a state-of-the-art model specialized for coding tasks, while \gpt is commonly used as a baseline in code generation. 

\subsection{Experimental Setup}
We implement an automated framework to run our experiments. It first runs tests on the buggy and correct GHRB repositories and records which tests fail or pass to confirm expected behavior. It then removes the test oracle from the bug-revealing test in both versions and appends the oracleless test, along with other relevant information, to the LLM input. The framework inserts the LLM-generated oracles into both versions, compiles, and runs the tests. It checks whether tests compile and which tests pass or fail, comparing results to the original output to assess accuracy and correctness. Each experiment is repeated five times to account for variability in LLM outputs.

This controlled configuration simulates a workflow in which an automated test generator (e.g., \textsc{EvoSuite}~\cite{fraser2011evosuite}) produces numerous assertion-free tests. The assertions typically generated by \textsc{EvoSuite} are regression oracles that replicate existing behavior and are therefore not suitable for detecting faults in the current version. In our setup, the LLM complements such tools by generating new, potentially fault-revealing assertions. Similar hybrid pipelines have been explored in prior work, such as \textsc{TOGLL}~\cite{binta2024togll}. Hence, although our setup is artificial, it closely mirrors a plausible integration of LLMs within modern automated testing workflows.

Each experiment combines an LLM, prompting technique, and prompt content. Since \starcoder is code-only~\cite{li2023starcoder}, it supports zero-shot and few-shot prompting. \gpt, with NLP capabilities, also supports CoT and ToT~\cite{huggingface2025}. This results in \textbf{$\mathbf{2{,}160}$ total runs} = $36$ bugs × $3$ prompt contexts × $(2 + 4)$ techniques × $5$ repetitions.

In all experiments, we set the LLM \texttt{temperature} to 0 and \texttt{top-p} to 1, following prior studies~\cite{schafer2023empirical,siddiq2023exploring,lops2024system}. Indeed, lower temperatures are often preferable when generating code~\cite{openai2023gpt4}.

\subsection{Metrics}
We evaluated the accuracy of the generated test oracles by comparing the outputs of tests with the replaced oracles to those with the original oracles on both the buggy and correct versions~\cite{watson2020learning}. To assess correctness, we measured the following:

\textbf{- Accuracy:} Tests correctly differentiating between buggy and correct versions: The number of cases in which the oracle(s) demonstrated consistent behaviour by correctly failing for the buggy version and passing for the correct version of the code. This is the optimum result.

\textbf{- Buggy Accuracy:} Tests correctly failing on buggy version: The number of test cases where the generated oracle(s) correctly identified a failure in the buggy version of the code.

\textbf{- Correct Accuracy:} Tests correctly passing on correct version: The number of test cases where the generated oracle(s) correctly passed the relevant tests in the correct version of the code.

\textbf{- Compilation Rate:} The rate at which generated test oracles compile.


These metrics evaluate the accuracy and performance of LLM-generated oracles, enabling a comparative analysis across various prompt configurations.

\section{Results}

\subsection{RQ1 - Input Context}
RQ1 : \emph{What influence does the content
and context within the prompt have on the accuracy of LLM generated test oracles?}

Table~\ref{tab:prompt_content_comparison} presents the average accuracy across all experiments for all 36 bugs for each prompt content scenario studied. We report the average across five runs, as the observed variation between runs was minimal.
The results indicate that providing more context in the prompt improves compilation rates. Specifically, prompts containing only the test prefix achieve an average compilation rate of 49.36\%, while adding the MUT increases this rate to 59.23\%, and incorporating the CUT raises it further to 76.26\%. 

Table~\ref{tab:prompt_content_comparison} also shows that the LLM-generated oracles cause buggy versions to fail correctly more often than they enable correct versions to pass correctly. For instance, with just the test prefix, buggy versions fail correctly 47.78\% of the time, compared to passing 42.22\% of the time for correct versions. When the MUT is included, these rates increase to 53.08\% and 43.85\%, respectively. Incorporating the CUT results in further improvements, with buggy versions failing correctly at 70.84\% and correct versions passing correctly at 58.13\%. These findings suggest that while LLMs can generate bug-revealing oracles, the generated oracles may still incorrectly fail on the correct versions of the code, indicating that  some identified bugs may not align with the expected program behaviour.

The results show that adding the CUT yields a greater improvement than adding the MUT. Transitioning from the test prefix to the MUT increases the buggy version failure rate by 5.30\% and the correct version pass rate by 1.63\%. However, moving from the MUT to the CUT leads to much larger gains: a 17.76\% increase in the buggy failure rate and a 14.28\% increase in the correct pass rate. These findings indicate that while the MUT offers some useful context, the CUT provides far more critical information, greatly enhancing the LLM's ability to generate accurate test oracles.

When considering overall accuracy, test oracles generated with only the test prefix achieve 40.74\%, which slightly decreases to 40.38\% when the MUT is added. However, including the CUT leads to a significant increase, with accuracy reaching 53.64\%. The negligible change between the test prefix and MUT scenarios, compared to the substantial improvement when introducing the CUT, reinforces the conclusion that greater contextual information in prompts enhances the LLM's ability to generate correct and accurate test oracles. This result suggests that the absence of CUT context may hinder the LLM's understanding of the code under test, reducing the quality of the generated oracles.

\begin{custombox}{Answering RQ1}
LLMs generate compilable and accurate test oracles more consistently when the prompt includes the CUT.
\end{custombox}

\begin{table}[t]
\rowcolors{1}{}{gray!10}
\caption{\textsc{Prompt content comparison (RQ1)}}

\centering
\setlength{\tabcolsep}{3pt}
\renewcommand{\arraystretch}{1}
\resizebox{\linewidth}{!}{
\begin{tabular}{l|c|c|c|c}
	\hiderowcolors

\toprule
\textbf{Content} & \textbf{Acc.\%} & \textbf{Buggy Acc.\%} & \textbf{Correct Acc.\%} & \textbf{Comp. Rate\%} \\
\midrule
\showrowcolors

Test Prefix       & 40.74 & 47.78 & 42.22 & 49.36 \\
Prefix + MUT & 40.38 & 53.08 & 43.85 & 59.23 \\
Prefix + CUT & 53.64 & 70.84 & 58.13 & 76.26 \\ \bottomrule
\end{tabular}
}
\label{tab:prompt_content_comparison}
\end{table}

\subsection{RQ2 - Prompt Engineering}

RQ2 : \emph{How do different prompt engineering techniques impact the accuracy of LLM generated test oracles relative to each other?}

\begin{table}[t]
\rowcolors{1}{}{gray!10}

\caption{Prompting Technique Comparison (RQ2)}
\setlength{\tabcolsep}{3pt}
\centering
  \resizebox{\linewidth}{!}{%
\begin{tabular}{l|c|c|c|c}
	\hiderowcolors
  \toprule
\textbf{Technique} & \textbf{Acc.\%} & \textbf{Buggy Acc.\%} & \textbf{Correct Acc.\%} & \textbf{Comp. Rate\%} \\ 
\midrule
\showrowcolors
Zero-Shot   & 54.56 & 64.47 & 57.28 & 67.38 \\ 
Few-Shot    & 51.30 & 68.33 & 55.00 & 72.96 \\ 
CoT         & 31.11 & 38.52 & 32.59 & 44.44 \\ 
ToT         & 29.26 & 41.11 & 32.22 & 44.81 \\ \bottomrule
\end{tabular}
\label{tab:prompting_technique_comparison}
}
\end{table}

Table~\ref{tab:prompting_technique_comparison} presents the average accuracy across all bugs for each prompting technique studied. We report the average across five runs, as the observed variation between runs was minimal. The results demonstrate that few-shot and zero-shot prompting techniques produce LLM-generated oracles with higher compilation rates compared to CoT and ToT techniques. Specifically, zero-shot scenarios achieve a compilation rate of 67.38\%, few-shot scenarios reach 72.96\%, while CoT and ToT techniques result in lower compilation rates of 44.44\% and 44.81\%, respectively.

The data also reveal a significant increase in the rate of buggy versions failing correctly compared to correct versions passing correctly across all prompting techniques. For zero-shot prompting, the rate increases from 57.28\% for correct versions passing to 64.47\% for buggy versions failing correctly. Similarly, for few-shot prompting, the rate rises from 55.00\% to 68.33\%. CoT and ToT prompting exhibit smaller but still notable increases with CoT going from 32.59\% to 38.52\% and ToT from 32.22\% to 41.11\%. These results, consistent with the findings from \textit{RQ1}, suggest that LLM-generated test oracles may not fully align with the expected behaviour, leading to failures on both buggy and correct versions of the code.

Zero-shot prompting achieves the highest accuracy at 54.56\%, followed by few-shot prompting at 51.30\%. CoT and ToT show significantly lower accuracies at 31.11\% and 29.26\%, respectively. The slight drop from zero-shot to few-shot may result from examples in few-shot prompts leading the LLM to deviate from expected behavior. The low performance of CoT and ToT may stem from their broader reasoning scope, which can cause the LLM to generate oracles that address general scenarios rather than the specific test prefix.

However, when just considering compiling oracles, zero-shot scenarios have an accuracy of 80.97\%, few-shot scenarios with 70.31\%, CoT scenarios with 70.00\%, and ToT scenarios with 65.30\%. This shows, when generating oracles that compile, while still demonstrating worse accuracy than zero-shot and few-shot scenarios, CoT and ToT scenarios are not significantly less accurate as the raw accuracies would suggest. Therefore, alongside the broadened scope for CoT and ToT prompts, a major issue is ensuring the LLM produces compilable oracles using these prompting techniques.

\begin{custombox}{Answering RQ2}
LLMs more consistently generate compilable and accurate test oracles with zero-shot and few-shot prompting technique compared to reasoning based techniques like CoT and ToT.
\end{custombox}

\subsection{RQ3 - Model Comparison}
RQ3 : \emph{ How do different LLMs impact the accuracy of LLM generated test oracles? }

Table~\ref{tab:rq3} shows the average results for each LLM using zero-shot and few-shot prompts. \starcoder achieves a much higher average compilation rate than \gpt, which is expected given \starcoder's focus on code generation~\cite{li2023starcoder}, unlike the more general-purpose \gpt. This limits \gpt's performance, with an average accuracy of 49.63\% compared to \starcoder's 56.31\%. \starcoder also outperforms \gpt in both average buggy and correct accuracy for these prompting techniques.

Table~\ref{tab:rq3-all} includes the CoT and ToT scenarios used with \gpt which could not be used with \starcoder. Here, we see the CoT and ToT prompts significantly decrease \gpt's average compilation rate and accuracies – \gpt's average compilation rate and accuracy are reduced to 51.02\% and 39.54\% respectively.

\begin{table}[t]
\caption{LLM Comparison – Zero-Shot and Few-Shot (RQ3)}
\label{tab:rq3}
\rowcolors{1}{}{gray!10}
\setlength{\tabcolsep}{3pt}
\centering
\resizebox{\linewidth}{!}{
\begin{tabular}{l|c|c|c|c}
\toprule
\showrowcolors
\textbf{LLM} & \textbf{Acc\%} & \textbf{Buggy Acc.\%} & \textbf{Correct Acc.\%} & \textbf{Comp. Rate\%} \\ 
\midrule
\starcoder & 56.31 & 79.42 & 61.36 & 83.69 \\ 
\gpt & 49.63 & 54.07 & 51.11 & 57.41 \\ \bottomrule
\end{tabular}
}
\end{table}

\begin{table}[t]
\caption{LLM Comparison – All Prompts (RQ3)}

\label{tab:rq3-all}
\rowcolors{1}{}{gray!10}

\setlength{\tabcolsep}{3pt}
\centering
\resizebox{\linewidth}{!}{
\begin{tabular}{l|c|c|c|c}
\toprule
\textbf{LLM} & \textbf{Acc.\%} & \textbf{Buggy Acc.\%} & \textbf{Correct Acc.\%} & \textbf{Comp. Rate\%} \\ 
\midrule
\starcoder & 56.31 & 79.42 & 61.36 & 83.69 \\ 
\gpt & 39.54 & 46.94 & 41.76 & 51.02 \\ \bottomrule
\end{tabular}
}
\end{table}

However, when considering only test oracles that successfully compile, \gpt demonstrates a competitive capacity for generating accurate oracles. Specifically, the accuracy of \gpt's compiling test oracles is 86.45\% compared to \starcoder's 67.28\% when considering zero-shot and few-shot scenarios. Even when including compiling oracles from CoT and ToT prompts, \gpt achieves an accuracy of 77.50\%. Though lower than zero-shot and few-shot accuracy, this shows \gpt's potential for reliable oracle generation in specific conditions.

\medskip
\begin{custombox}{Answering RQ3}
\starcoder-generated test oracles demonstrate a higher average compilation rate than \gpt generated ones resulting in higher accuracies.
\end{custombox}

\begin{table}[t]
\caption{Average results for each configuration (RQ4)}
\label{tab:rq4}
		\rowcolors{1}{}{gray!10}

\centering
\resizebox{\linewidth}{!}{
\begin{tabular}{l|c|c|c|c|c}
\toprule
 \textbf{Config} & \textbf{Acc. Rank} & \textbf{Acc.\%} & \textbf{Buggy Acc.\%} & \textbf{Correct Acc.\%} & \textbf{Comp. Rate\%} \\
 \midrule
S.Z.T.  & 4  & 66.67 & 77.78 & 66.67 & 77.78 \\
S.Z.M.  & 3  & 68.57 & 84.29 & 71.43 & 88.57 \\
S.Z.C.  & 6  & 55.29 & 88.24 & 60.00 & 92.94 \\
S.F.T.  & 5  & 55.56 & 66.67 & 55.56 & 66.67 \\
S.F.M.  & 9  & 44.44 & 77.78 & 55.56 & 83.33 \\
S.F.C.  & 7  & 50.00 & 83.33 & 61.11 & 94.44 \\
G.Z.T.  & 18 & 22.22 & 22.22 & 31.11 & 31.11 \\
G.Z.M.  & 10 & 44.44 & 44.44 & 44.44 & 44.44 \\
G.Z.C.  & 2  & 73.33 & 75.56 & 73.33 & 75.56 \\
G.F.T.  & 12 & 37.78 & 44.44 & 37.78 & 44.44 \\
G.F.M.  & 8  & 45.56 & 51.11 & 45.56 & 62.22 \\
G.F.C.  & 1  & 74.44 & 86.67 & 74.44 & 86.67 \\
G.Ch.T. & 15 & 26.67 & 33.33 & 26.67 & 33.33 \\
G.Ch.M. & 17 & 22.22 & 28.89 & 28.89 & 44.44 \\
G.Ch.C. & 11 & 40.00 & 53.33 & 42.22 & 55.56 \\
G.Tr.T. & 13 & 35.56 & 42.22 & 35.56 & 42.22 \\
G.Tr.M. & 16 & 23.33 & 38.89 & 23.33 & 38.89 \\
G.Tr.C. & 14 & 28.89 & 42.22 & 37.78 & 53.33 \\ \bottomrule
\end{tabular}
}
\end{table}

\begin{figure*}[t]

\centering
\includegraphics[width=17cm]{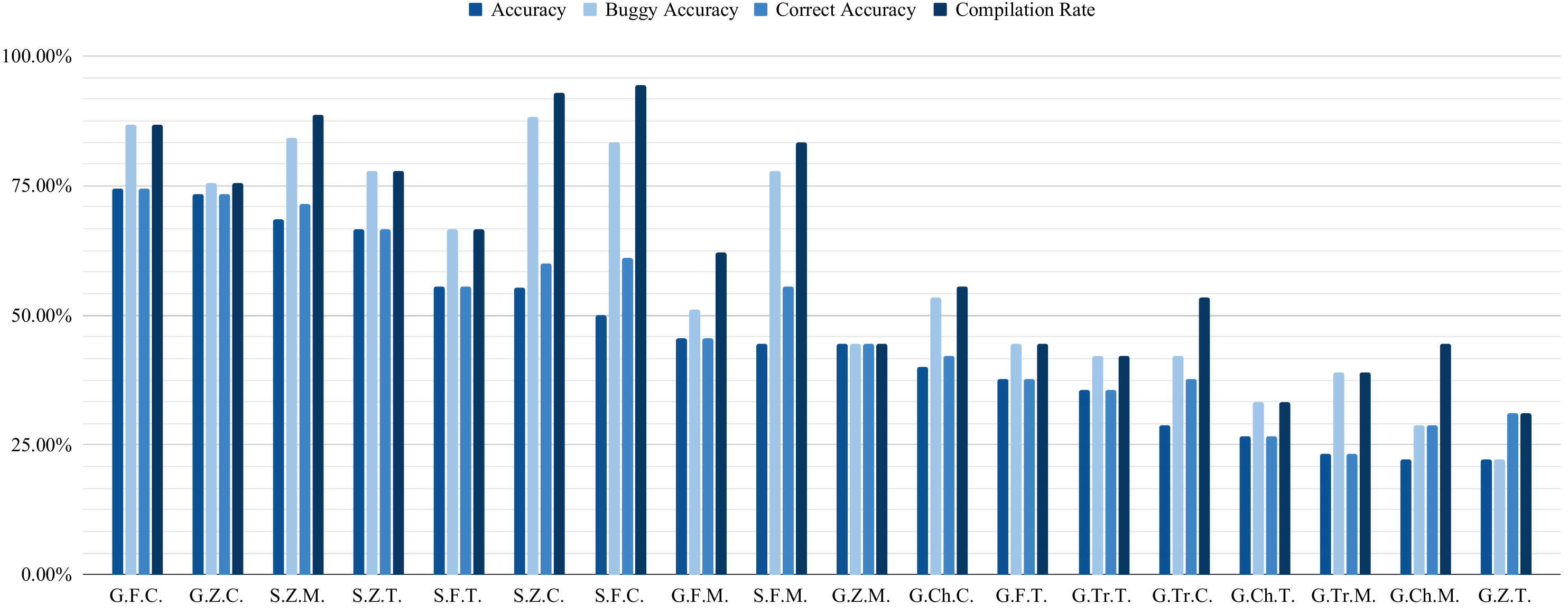}
\caption{Bar chart of results from all experiments ordered by average accuracy grouped by configuration (RQ4)}

\label{graph}
\end{figure*}

\subsection{RQ4 -Impact Analysis}
RQ4 : \emph{Which factor has the most significant impact on LLM generated test oracles?}

Table~\ref{tab:prompt_content_comparison} shows that including the test prefix and the CUT has the highest average accuracy of 53.64\% and including the test prefix and the MUT has the lowest average accuracy of 40.38\%, giving a range of 13.26\%. Considering the prompting technique, from Table~\ref{tab:prompting_technique_comparison}, zero-shot has the the highest accuracy of 54.56\% and ToT has the lowest accuracy of 29.26\%, giving a range of 25.30\%. Considering the LLM used, from Table~\ref{tab:rq3-all}, \starcoder has the highest accuracy of 56.31\% and \gpt has the lowest accuracy of 49.63\% for cases where the same prompting techniques are used, giving a range of 6.68\%.  These findings indicate the prompting technique is the most significant factor out of these because of the far greater range of results observed compared to the content and LLM. This is followed by the prompt content, and then the LLM.

Table~\ref{tab:rq4} ranks each scenario's average accuracy. Column "Config" represents the \texttt{LLM.} \texttt{Prompt-Strategy.} \texttt{InputContent} setup (e.g., \textit{G.Ch.C.} denotes \textit{GPT, CoT, and Text Prefix + CUT}). We have introduced the Labels when describing the components individually (e.g., Text Prefix + CUT (\textbf{C}) in Section II).

Column ``Acc. Rank'' gives the ranking of the configurations ordered by Column `Acc.\%'. 
From this column we can calculate average rank of each different component. Out of the 18 scenarios conducted, the average rank of \starcoder scenarios is 5.67 and the average rank of \gpt scenarios is 11.42. The average rank of zero-shot  is 7.17, few-shot  is 7.00, CoT  is 14.33, and ToT scenarios is 14.33 too. The average rank for just test prefix scenarios is 11.17, for test prefix and MUT scenarios is 10.5, and for test prefix and CUT scenarios is 6.83. Because \starcoder is not tested with CoT and ToT techniques, if we remove these six scenarios, the average rank of \starcoder is 5.67, \gpt is 7.33, zero-shot scenarios is 6.17, few-shot scenarios is 6.83, test prefix scenarios is 8, test prefix and MUT scenarios is 7.5, and test prefix and CUT scenarios is 4.00.

Figure~\ref{graph} ranks the data from Table~\ref{tab:rq4} in descending order by average accuracy. From this, we compute the average rank of each factor across the 18 combinations. \starcoder scenarios rank highest with an average of 5.67, followed by prompts with the CUT at 6.83, and few-shot prompts at 7. In contrast, CoT and ToT have the lowest average rank at 14.33. These results suggest that prompting technique is the most influential factor, as it shows the greatest variation in performance across the rankings.

We expect \starcoder to rank among the highest in accuracy due to its code-focused training and exclusion from CoT/ToT prompts. Without these scenarios, \starcoder still outperforms \gpt (5.67 vs. 7.33), but CUT scenarios now have the highest rank (4), highlighting their importance for generating accurate test oracles.

\begin{custombox}{Answering RQ4}
The prompting technique has the greatest impact on the accuracy of LLM-generated test oracles.
\end{custombox}

\section{Discussion}
\subsection{Improvements with the CUT (RQ1)}
The findings for \textit{RQ1} show that including the CUT significantly improves both accuracy and compilation rates of LLM-generated oracles. While adding the MUT boosts compilation over the test prefix alone, likely due to added method context, it results in lower overall accuracy. This suggests the MUT provides limited guidance, whereas the CUT offers essential functional context and usage patterns for accurate oracle generation.


\subsection{Prompting Strategies (RQ2)}

For \textit{RQ2}, it is evident that zero-shot and few-shot prompting scenarios outperform CoT and ToT strategies across all metrics. We attribute this to CoT and ToT encouraging broader reasoning, which often seem to lead to oracles that include irrelevant or unintended behavior. This findings suggest that while useful for general reasoning~\cite{wei2022chain}, these strategies may hinder concise, behavior-specific oracle generation. Future work could combine CoT/ToT with few-shot examples aligned with expected behaviour to balance reasoning strength with improved accuracy and compilation.



\subsection{Model-Specific Performance (RQ3)}

The comparative analysis of \starcoder and \gpt in zero-shot and few-shot scenarios reveals key differences. \starcoder, trained exclusively on code, achieves higher accuracy and compilation rates than \gpt, producing compilable code more consistently~\cite{xiong2023program}. However, \gpt better reflects expected program behaviour despite a lower compilation rate. Addressing these contrasting strengths and weaknesses could involve combining insights from both models to design prompting techniques or hybrid approaches that leverage \starcoder's compilation strength and \gpt's capacity for accurate behaviour representation.


\subsection{Other Considerations (RQ4)}
The findings from \textit{RQ4} show that using \starcoder gives the most accurate test oracles overall. However, \starcoder was not tested with CoT and ToT, which lowers the average accuracy of the other scenarios since CoT and ToT perform poorly. When we remove CoT and ToT, we see that adding the CUT has a bigger positive effect on accuracy than using \starcoder.

From Figure~\ref{graph}, we see that the highest average accuracies come from using \gpt with the CUT. \textit{RQ3} also shows that for compiling oracles, \gpt has higher average accuracy than \starcoder. This is clear in Figure~\ref{graph}, especially in the S.Z.C., S.F.C., and S.F.M. cases, where the oracles compile well but have lower accuracy. 
With more context, \gpt generates more accurate oracles than \starcoder, but with less context, \starcoder performs better. Thus, the best LLM depends on the prompt's level of context.

Although the evaluation scenario is controlled, it remains useful for understanding how prompting and context influence oracle synthesis. We assume the faulty region is already localized and the test prefix exercises that region. This abstraction removes confounding factors such as bug localization or test generation, focusing instead on the LLM’s ability to complete a potentially fault-revealing test.

\subsection{Further LLM Output Analysis}
We manually examined a subset of eight test oracle generations from \starcoder and eight from \gpt to gain further insights into their behaviour.

For \starcoder, cases where oracles correctly identified failures in buggy code but incorrectly flagged correct code tended to fall into two main categories: over generation of oracles~\cite{terragni-fse-2020} and insufficient input context. Regarding over generation, \starcoder often produced more oracles than necessary. While this set might include the correct oracles, it also contains irrelevant ones that fail due to the limited context provided by the test prefix.
For example, it was common to see oracles testing multiple unrelated functionalities. This limitation likely results from the narrow scope of the input prompt used in our experiments. Supplying broader context, such as the entire test class or suite, could help the model assign oracles to the correct test cases and improve their relevance. The second issue relates to the lack of context in the input itself. When only the test prefix is provided, \starcoder lacks sufficient information to generate reliable oracles. Without enough details about the code’s purpose and usage, the model struggles to create representative oracles. 
Figure~\ref{fig:lackcontext} shows a \starcoder-generated oracle failing to compile due to a missing \texttt{isSafe} method.

\begin{figure}[t]
    \centering
    \begin{minipage}{\linewidth}
      \begin{lstlisting}[language=Java]
public void testCopyConstructor_noSideEffectOnAttributes() {
    Safelist safelist1 = Safelist.none().addAttributes(TEST_TAG, TEST_ATTRIBUTE);
    Safelist safelist2 = new Safelist(safelist1);
    safelist1.addAttributes(TEST_TAG, "invalidAttribute");

    (*@\textcolor{violet}{$\backslash\backslash$ Original Oracle\\
assertFalse(safelist2.isSafeAttribute(TEST\_TAG, null, new Attribute("invalidAttribute", TEST\_VALUE)));}@*)
    (*@\textcolor{blue}{$\backslash\backslash$ Generated Oracle\\
    assertFalse(safelist2.isSafeTag(TEST\_TAG));}@*)
}
\end{lstlisting}
    \end{minipage}

    \caption{Example of LLM-generated oracle  Lacking Context }
    \label{fig:lackcontext}
\end{figure}





GPT-4o, often failed correct code for similar reasons to \starcoder, but also frequently overcomplicated comparisons. Figure~\ref{fig:overcompl} is an example where instead of using \texttt{assertEquals} to compare two values, it would use \texttt{assertArrayEquals}. 
This likely stems from \gpt trying to compare entire objects rather than just method results, aiming for content-based comparison.
However, this can leads to both false positive and false negative results as they are comparing objects which were not originally intended to be compared as a part of the particular test case. For instance, the constructor and overridden \texttt{equals} method of the objects could mean the two instances of the object being compared are considered equal despite the values the test actually wants to compare being different, thus potentially resulting in a false positive. Conversely, if the \texttt{equals} method for the objects being compared has not been overridden or uses object references to compare, the two instances being compared (if they are different instances) will be returned as not equal, thus resulting in a false negative.



Both \starcoder and \gpt struggled with generating oracles that failed to compile, mainly due to missing context and over generation of code. When given only the test prefix, the models lacked details about methods, arguments, or expected outputs, leading to guesswork and compilation errors. Including the MUT helped but didn’t fully resolve the issue, as it provided context for only one method. Without including the CUT, aligning output with the full test scenario remained difficult. Over generation of code was especially the case for Chain-of-Thought and Tree-of-Thought prompts, where the models often added unnecessary test logic. 

To address these limitations, future work could explore providing more comprehensive prompts, such as including the test class or suite, to give LLMs a clearer understanding of the broader context. This might enable the models to allocate oracles to the appropriate test cases and reduce over generation of irrelevant test logic. Additionally, ensuring that prompts guide LLMs toward simpler, more direct assertions could help avoid overcomplicated oracle logic and improve both the accuracy and reliability of generated test oracles.

\begin{figure}[t]
    \centering
    \begin{minipage}{\linewidth}
\begin{lstlisting}[language=Java]
public void testLazilyParsedNumberDeserialization() {
    LazilyParsedNumber expected = new LazilyParsedNumber("1.5");
    LazilyParsedNumber actual = gson.fromJson("1.5", LazilyParsedNumber.class);

    (*@\textcolor{violet}{$\backslash\backslash$ Original Oracle\\
    assertEquals(expected, actual);}@*)
    (*@\textcolor{blue}{$\backslash\backslash$ Generated Oracle\\
    assertArrayEquals(new LazilyParsedNumber[]{expected}, new LazilyParsedNumber[]{actual});}@*)
}
\end{lstlisting}
    \end{minipage}

    \caption{Example of overcomplicated LLM-Generated Oracle}
    \label{fig:overcompl}
\end{figure}

\section{Threats to Validity}

\noindent \textbf{Data leakage.} To reduce the risk of data leakage, we use the GHRB dataset~\cite{lee2024github}, which includes bugs reported after \starcoder and \gpt training cutoff dates, ensuring the models have not seen these faults.

\noindent \textbf{Sample size and generalisation.} We used 36 randomly selected bugs from GHRB. The nature of these bugs may affect our findings, and results may not generalise beyond this sample. 

\noindent \textbf{Bug-revealing test case prefixes.} Our study uses test case prefixes known to expose bugs. This limits our ability to assess whether LLMs can generate effective oracles without such prefixes or in correct code scenarios.

\noindent \textbf{Lack of false positives analysis.} We evaluate only buggy scenarios. However, in practice, most code is correct. It is also important to check whether LLMs generate false positives: assertions that wrongly fail on correct code~\cite{dinella2022toga,liu2023realisticevaluationneuraltest}.  Evaluating oracle generation for bug-free units is  an important direction for future work.

\noindent \textbf{Randomness in LLM outputs.} To account for the non-deterministic nature of LLMs~\cite{cho2025metamorphic}, we ran prompts with each parameter combination five times. Our results indicate that the variance across these runs was low. Additionally, we used fixed values for temperature and top-p parameters. We acknowledge that exploring a broader range of values for these parameters may affect results and consider this an avenue for future work.


\noindent \textbf{Prompt configuration scope}. A potential threat to validity lies in the limited range of input prompt configurations explored. We only evaluated three scenarios (test prefix alone, test prefix with MUT, and test prefix with CUT), while excluding other possible variations such as removing comments, including dependencies, or using only method and class signatures. These alternative configurations may influence LLM behaviour and the resulting oracle generation in ways we did not capture. As such, our findings may not fully generalise across different prompt structures.

\noindent \textbf{Scope of LLM selection.} Finally, our study evaluates only one representative LLM from each category—one code-specific (\starcoder) and one general-purpose (\gpt). This choice was made for feasibility reasons and because the GHRB dataset is specifically designed to address data leakage concerns related to these models. \emph{Future studies should explore a broader range of LLMs to assess the generality of our findings}.

\section{Related Work}


Automated unit test and oracle generation is a research area that has gained significant attention~\cite{jahangirova2023sbfttrack,terragni-ase-2018}. Tools like \textsc{EvoSuite}~\cite{fraser2011evosuite} and \textsc{Randoop}~\cite{pacheco2007randoop} generate tests with regressions assertions, while recent methods use neural models to generate assertions from test prefixes~\cite{watson2020learning, yu2022automated, tufano2022generating, dinella2022toga, hossain2023neural, shin2024assessing}. Due to space constraints, in this section we highlight only key studies on LLM-based test and oracle generation.

\medskip
\noindent
\textbf{LLMs for Test Case/Suite Generation. }
Recent studies have explored LLMs for automated test generation, focusing on their effectiveness, prompt design, and limitations. Yang et al.~\cite{yang2024evaluation} found that LLM-generated unit tests had lower compilation rates and coverage compared to \textsc{EvoSuite}. Alshahwan et al.~\cite{alshahwan2024automated} introduced \textsc{TestGen-LLM}, deployed at Meta, which improved 11.5\% of test cases, with 73\% accepted in production. Siddiq et al.~\cite{siddiq2023exploring} analyzed \textsc{Codex}, \textsc{GPT-3.5}, and \starcoder, highlighting the impact of code context. Schäfer et al.~\cite{schafer2023empirical} evaluated the LLM-driven test tool \textsc{TestPilot}, in terms of code coverage and failure detection. Lops et al.~\cite{lops2024system} introduced \textsc{AgoneTest}, noting low compilation rates and the influence of prompting strategies. Ouédraogo et al.~\cite{ouedraogo2024llms} found that structured prompts improved test quality but struggled with complex code.

All of the mentioned studies focus on generating full test cases using LLMs. In contrast, our study addresses a more specific challenge: automated test oracle generation. This focus is well justified, as the oracle problem is one of the main bottlenecks in achieving full test automation~\cite{barr2014oracle}. By isolating test oracle generation, we remove confounding factors related to the test prefix (e.g., coverage, code quality) and focus solely on the test oracle's fault detection effectiveness. 

\medskip
\noindent
\textbf{LLMs for Test Oracle Generation}
Research related to test oracle generation with LLMs has advanced significantly, with various approaches exploring the potential of LLMs to automate and improve software testing processes. The work by Molina et al.~\cite{molina2024test} presents a
roadmap for future research on the usage of LLMs for test oracle automation. 
Hossain et al.~\cite{binta2024togll} introduced \textsc{TOGLL}, a method that leverages LLMs for generating test assertions while relying on the \textsc{EvoSuite} tool for test prefix generation.
The authors evaluate six different levels of contextual information.  
Hayet et al.~\cite{hayet2024c} introduce \textsc{ChatAssert}, an LLM-based test oracle generation tool with two modes of execution: generation and
repair. In generation mode, it uses ChatGPT with a fixed prompt that includes summaries of the methods used in the test prefix, as well as similar examples in the form of other tests from the same test file.
Konstantinou et al.~\cite{konstantinou2024llms} studied whether LLMs generate test oracles that focus on the implemented behaviour of the code or whether they are capable of producing non-regression oracles that capture the expected behaviour. The authors reuse the best-performing prompts from previous works such as TOGLL~\cite{binta2024togll} and \textsc{ChatTester}~\cite{yuan2023no}. 
Zhang et al.~\cite{zhangexploring} investigated the performance of LLM-based assertion generation in terms of bug detection. The prompts employed by the authors contain the test prefix and MUT.



%

Our study is the first of its kind, substantially differing from previous works. First, most prior studies evaluate a single prompt type, varying only the fixed information about the MUT or CUT. The exception is Hossain et al., who explored multiple prompts with different levels of MUT context. However, no study has examined the combination of both input and prompt strategies. Second, we are the first to use the GHRB dataset—designed to reduce data leakage—for oracle generation, addressing a key limitation in earlier Java-based studies.


\section{Conclusions and Future Work}

This study provides a baseline for LLM-driven fault-revealing test oracle generation, an essential step toward reliable Promptware. We evaluated two LLMs (\starcoder and \gpt) across four prompting techniques (zero-shot, few-shot, CoT, ToT) and four input contexts (test prefix, method under test, full class). 
Our large-scale evaluation (over 2,000 runs) highlights key challenges and insights that can help improve oracle generation in the FM era. Assertion compilability remains a major obstacle, and including full class context improves oracle quality by 12.9\%. Simpler prompting techniques outperform CoT and ToT by 23\%, and \starcoder consistently outperforms \gpt. Prompting technique emerges as the most impactful factor.

Future work should focus on improving assertion compilability, possibly through static analysis or LLM-driven refinement prompts~\cite{ravi2025llmloop}. Additionally, hybrid prompting strategies that combine structured reasoning (e.g., CoT) with simpler approaches (e.g., zero-shot or few-shot) may improve both accuracy and reliability.

\section
*{Acknowledgments}

This work has been supported by the ITEA grant GENIUS (project number 23026).

\bibliographystyle{IEEEtran}
\bibliography{bib}

\end{document}